\documentclass[journal,twoside,web]{ieeecolor}
\usepackage{generic}
\usepackage{cite}
\usepackage{amsmath,amssymb,amsfonts}
\usepackage{algorithmic}
\usepackage{graphicx}
\usepackage{algorithm,algorithmic}
\usepackage{hyperref}
\hypersetup{hidelinks=true}
\usepackage{textcomp}
\usepackage{booktabs}

\def\BibTeX{{\rm B\kern-.05em{\sc i\kern-.025em b}\kern-.08em
    T\kern-.1667em\lower.7ex\hbox{E}\kern-.125emX}}
\begin{document}
\title{Voluntary Triggering of Shared-Autonomous Prosthetic Control via IMU-Based Motion Gestures}
\author{Aabira Zaman, Kaijie Shi, \IEEEmembership{Student Member, IEEE}, and Xianta Jiang, \IEEEmembership{Senior Member, IEEE}
\thanks{Submitted on September 19th, 2025. This work was supported in part by the Government of Canada's New Frontiers in Research Fund (NFRF, Grant No NFRFE-2022-00407) and Natural Sciences and Engineering Research Council of Canada's Research Tools and Instruments (NSERC RTI, Grant No  RTI-2022-00688).}
\thanks{This work involved human subjects or animals in its research. Approval of all ethical and experimental procedures and protocols was granted by the Memorial University Interdisciplinary Committee on Ethics in Human Research (20210316-SC).}
\thanks{
Aabira M.A Zaman, Kaijie Shi are with Department of Computer Science, Memorial University of Newfoundland, St. John’s, NL A1B 3X5, Canada, (emails: amzaman@mun.ca,kaijies@mun.ca).
}
\thanks{
Xianta Jiang is with Department of Computer Science, Memorial University of Newfoundland, St. John’s, NL A1B 3X5, Canada. (email: xiantaj@mun.ca).
}
\thanks{*Corresponding author: Xianta Jiang (email: xiantaj@mun.ca)}
}

\maketitle

\begin{abstract}
Recently, a shared-autonomous scheme has been introduced into prosthetic hand control field, where the user provides high-level intent by moving the hand towards the target, and the artificial intelligence system autonomously executes low-level control (e.g., grasp and release the object). This system reduces user workload but risks unintended grasp or release actions without explicit user control. In particular, release actions remain challenging, as vision-based autonomous systems typically assume that proximity to a supporting surface signals the user’s intent to let go, making mid-air release tasks difficult and error-prone. This study presents an inertial measurement unit (IMU)-based gesture-triggered interface enabling voluntary initiation or override of grasp and release actions to the autonomous system. A real-time motion detection algorithm recognizes three deliberate upper-limb gestures: shoulder shrug, elbow flap, and wrist shake, across three control paradigms: autonomous, hybrid, and manual. In a controlled study with 14 able-bodied participants and one individual with an upper-limb difference, the elbow flap emerged as the most preferred gesture (66\% preference) and achieved 95\% mean successful rate. Manual mode produced the highest accuracy (95\%), while autonomous mode and hybrid mode were most preferred for daily use (38\%). Results suggest that IMU-based voluntary triggers enhance alignment between user intent and prosthetic action, improving reliability and perceived control. This approach offers a practical pathway toward safer, more adaptable prosthetic systems and can be extended to real-world applications requiring rapid, intentional overrides of autonomous behavior.
\end{abstract}

\begin{IEEEkeywords}
Inertial Measurement Unit (IMU), Prosthetic Control, Human-Machine Interaction, Voluntary Motion Trigger, Gesture Recognition, Motion Detection, Amputation, Assistive technology, Manual Control, Autonomous Control, Control interface, Algorithms.
\end{IEEEkeywords}

\section{Introduction}
\label{sec:introduction}
\IEEEPARstart{P}{rostheses} are vital to enhancing the quality of life for approximately 65 million amputees worldwide as of 2020, with an estimated 1.5 million amputations occurring annually~\cite{18, 26, 27}. To restore lost functionality, conventional upper-limb prosthetic systems rely on user-driven control methods such as surface electromyography (EMG)~\cite{8}, but EMG signals are often noisy~\cite{19,10}, fatigue-sensitive~\cite{10}, difficult to generalize and less viable for especially in the case of above-elbow  amputations amputees~\cite{4}. Alternative control methods such as voice commands are feasible but impose a social and privacy burden by requiring users to vocalize instructions~\cite{4, 13, 1}. These limitations underscore the need for more intuitive, robust, and accessible control alternatives.

Recent progress in robotics, sensing, and machine learning has enabled semi-autonomous prosthetic systems that combine computer vision with EMG-based intent detection. In such frameworks, visual input assists in determining grasp type, size, and orientation, while EMG cues indicate the timing of actions such as hand closure or release~\cite{21, 22, 23, 24, 25, 5}. Although this hybrid approach improves efficiency and responsiveness, existing implementations remain constrained by several factors: they typically rely on a limited set of predefined grip patterns, are optimized for only a narrow range of object types, and continue to depend heavily on muscle signals that are often inconsistent under real-world conditions.

It is believed that within a shared-control framework involving the human in the loop~\cite{Shared-survery, 20}, prosthetic hands can be made increasingly intelligent and autonomous to reduce both the physical and cognitive demands associated with operation. One research group has recently developed a bio-signal-free autonomous prosthetic hand control system capable of mimicking natural human manipulation, enabling adaptive grasping and releasing of objects without requiring explicit muscle-signal inputs from the user~\cite{kaijie}. Trained through imitation learning from human demonstrations, the system dynamically initiates and terminates grasps, adjusts hand posture, and regulates force in response to object context. Experiments demonstrate robust generalization to unseen objects, cluttered environments, and new users, achieving reliable human-like performance without bio-signal input or additional fine-tuning~\cite{kaijie}. 
However, the absence of user agency may result in unintended actions, such as incorrect grasps or releases, due to misinterpreted environmental cues or system misalignment with user intent~\cite{7}. This underscores the need for a voluntary trigger interface that allows users to consciously initiate or interrupt prosthetic actions in real time.

One promising approach is the use of motion-triggered control mechanism using a wearable Inertial Measurement Unit (IMU)~\cite{12}. These non-invasive sensors capture real-time acceleration and angular velocity, enabling reliable gesture-based prosthetic control that remains robust to muscle fatigue and environmental noise and consistent across users~\cite{1}. This study focuses on comparing multiple IMU-based motion control paradigms by evaluating their relative accuracy, comfort, and user preference to identify the most intuitive and reliable interaction strategy for prosthetic control. To facilitate this comparison, a motion detection algorithm was implemented to analyze deliberate upper-limb motion patterns from IMU data, allowing intentional gestures to initiate or terminate prosthetic actions on demand~\cite{8}.

Prior studies have shown the feasibility of IMU-assisted gesture recognition for assistive technologies \cite{1, 8, 11}. Building on this foundation, our work integrates gesture-based control into an autonomous prosthetic hand to enhance flexibility and user agency. Specifically, we design a real-time gesture detection algorithm, integrate it with an autonomous prosthetic system, and evaluate its performance and user acceptance across multiple gestures and control paradigms. We explore three intuitive triggering gestures: shoulder shrug, elbow flap, and wrist shake implemented through the proposed detection algorithm. 
Furthermore, the control interface is evaluated under three operating paradigms: 1) Autonomous Mode, a baseline mode in which the prosthetic functions entirely via its onboard intelligence, 2) Hybrid Mode, where the system performs grasping and the user triggers release, and 3) Manual Mode, where both grasp and release require gesture input.

In this work, we investigate whether IMU-based voluntary motion triggers can provide a practical and reliable mechanism for initiating or overriding prosthetic actions. The proposed framework was evaluated in a user study involving fourteen able-bodied participants and one individual with a congenital limb difference. Participants performed grasp-and-release tasks under three gesture types and three control paradigms, while performance, comfort, and subjective preference metrics were recorded. Results indicate that IMU-based voluntary triggers achieve high detection reliability and strong user acceptance, supporting their potential to enhance intuitive human–machine interaction and preserve user agency within intelligent prosthetic systems.

\section{Methodology and Materials}

To enable voluntary control within an autonomous prosthetic system, a wearable inertial measurement unit (IMU) was incorporated as a motion-based trigger interface. This IMU interface was integrated into the vision-based autonomous control system to establish a hybrid framework combining user intent with intelligent grasp automation.

\subsection{System Architecture}
\label{system_arch}
The proposed control system employs a multi-process software architecture integrating three primary hardware components: a prosthetic hand, a wrist-mounted camera, and a Bluetooth Low Energy (BLE) IMU sensor. The prosthetic hand (Ability Hand, Psyonic Inc.) executes grasp and release commands delivered by the control software, as shown in Fig.~\ref{fig:sys_arch}. A wrist-mounted RGB-D camera provides continuous visual feedback for object perception, while the IMU captures upper-limb motion signals to generate user-initiated commands.

For the control software, as shown in Fig.~\ref{fig:sys_arch}, six concurrent processes operate in parallel to ensure real-time control:
(1) a camera process capturing live visual input~\cite{kaijie}, (2) a hand control process managing bidirectional communication with the prosthesis~\cite{kaijie}, (3) a neural network process performing real-time action prediction~\cite{kaijie}, (4) an IMU process establishing BLE communication with IMU device for reading and processing motion signals (5) a gesture-detection process interpreting the processed IMU signals for voluntary triggers, and (6) a control-flow process coordinating the decision logic between autonomous and user-driven control streams.
Shared global variables synchronize hand state, force sensor feedback, visual frames, and gesture-detection events across processes using inter-process queues and manager dictionaries to ensure deterministic timing and data integrity.

\subsection{Hardware Setup}
\subsubsection{Inertial Measurement Unit (IMU)}
The system uses the WT901BLECL IMU (WitMotion Inc., Shenzhen, China). This sensor provides real-time measurements across nine degrees of freedom (3-axis acceleration, angular velocity, and magnetic field), along with angle, quaternion, and port status data. It supports configurable output rates up to 200 Hz and transmits data via BLE with a stable range upto 50 meters, making it well-suited for wearable applications and motion recognition tasks. 
In this system, accelerometer signals are processed for gesture-trigger detection, as shown in Fig.~\ref{fig:sys_arch}.

\subsubsection{Prosthetic Hand}
The system uses the Psyonic Ability Hand, a 6-DOF robotic prosthetic with one motor per finger (the thumb has 2 DOF). Each fingertip contains six force-sensitive resistors (FSRs), totaling 30 tactile sensors, capable of streaming high-frequency force and encoder data via USB or Bluetooth. It provides the current hand state information, e.g., joint positions and force values (FSRs) to the system and receives the target action position from the system. The device streams joint states and force sensor values while receiving target position commands from the control system (See Fig.~\ref{fig:sys_arch}).

\subsubsection{Wrist-Mounted Camera}
\begin{figure*}[!t]
    \centering    \includegraphics[width=\textwidth]{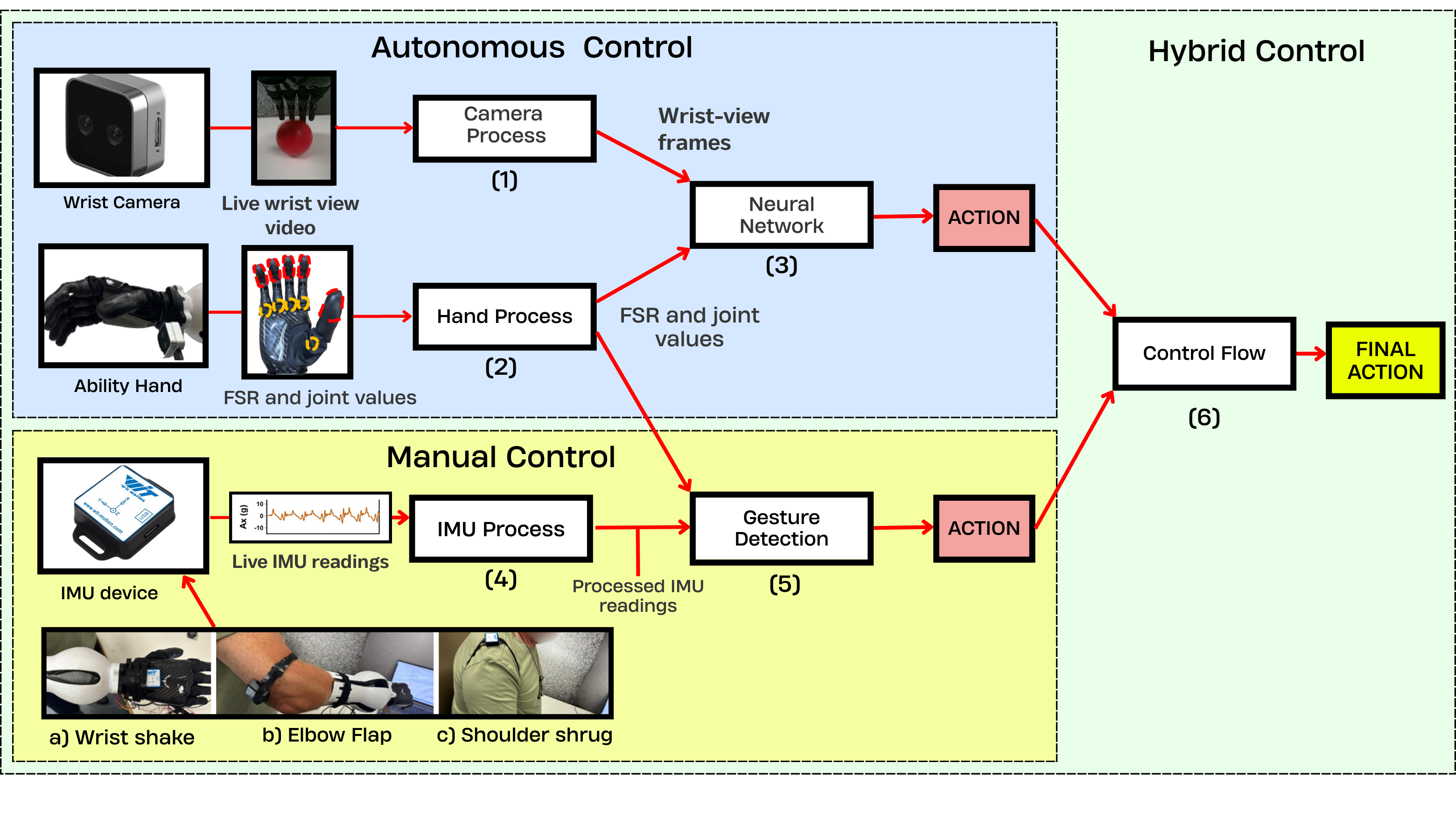}
    \caption{System architecture of the proposed prosthetic control framework. The diagram illustrates the integration of hardware components including the PSYONIC Ability Hand, wrist-mounted Intel RealSense camera, and WitMotion IMU to read the any one of the three candidate trigger gestures: (a) wrist-shake, (b) elbow-flap, (c) shoulder-shrug|and the six concurrent software processes: (1) camera process for live wrist view video, (2) hand control process for communication with the prosthetic device, (3)  neural-network process for predicting next action, (4) IMU process reading and processing live IMU data (5) Gesture Detection process to detect any voluntary gestures in the processed IMU motion readings, and (6) Control-flow process coordinating the autonomous system and the manual system to decide the final action to be either grab or release.}
    \label{fig:sys_arch}
\end{figure*}

A wrist-mounted RGB camera with a resolution of $320 \times 240$ pixels provides continuous visual input of the prosthetic hand and its surrounding environment, enabling real-time grasp prediction during autonomous operation. Its integration within the overall control framework is illustrated in Fig.~\ref{fig:sys_arch}.

\subsection{Autonomous Control} 
The baseline autonomous controller was developed by Shi et al.~\cite{kaijie}. It employs a vision-based neural network trained using an imitation-learning framework derived from human teleoperation demonstrations. Using synchronized visual input from a wrist-view camera along with proprioceptive and tactile inputs from the prosthetic hand, the model predicts continuous joint-position trajectories to achieve natural, human-like grasping and releasing motions. 
The wrist-view images are streamed to the neural-network process through a dedicated camera process, which continuously acquires live footage from the wrist-mounted camera. In parallel, a hand control process manages bidirectional communication with the prosthetic hand, providing the neural network with real-time tactile and joint-state data while executing the action commands generated by the network.
The autonomous control architecture is illustrated in Fig.~\ref{fig:sys_arch}. This fully autonomous pipeline operates continuously under normal conditions and is only interrupted when a voluntary IMU trigger temporarily pauses neural-network control.




\subsection{IMU-Based Manual Control} 
An IMU-based manual control system was developed to enable voluntary user-initiated operation of the prosthetic hand.
This system employs an IMU to detect deliberate upper-limb motions that serve as trigger signals for initiating prosthetic actions.
A custom real-time gesture detection algorithm was developed to recognize distinct upper-limb motion patterns corresponding to predefined trigger gestures. Once a valid gesture is identified, the system interprets it as an intentional command to either grasp or release.

\subsubsection{Upper-limb Gesture Triggers} 
Three deliberate upper-limb motions were selected as candidate trigger gestures: wrist shake, elbow flap, and shoulder shrug. 

\begin{figure}[htbp]
    \centering
    \includegraphics[width=1\linewidth]{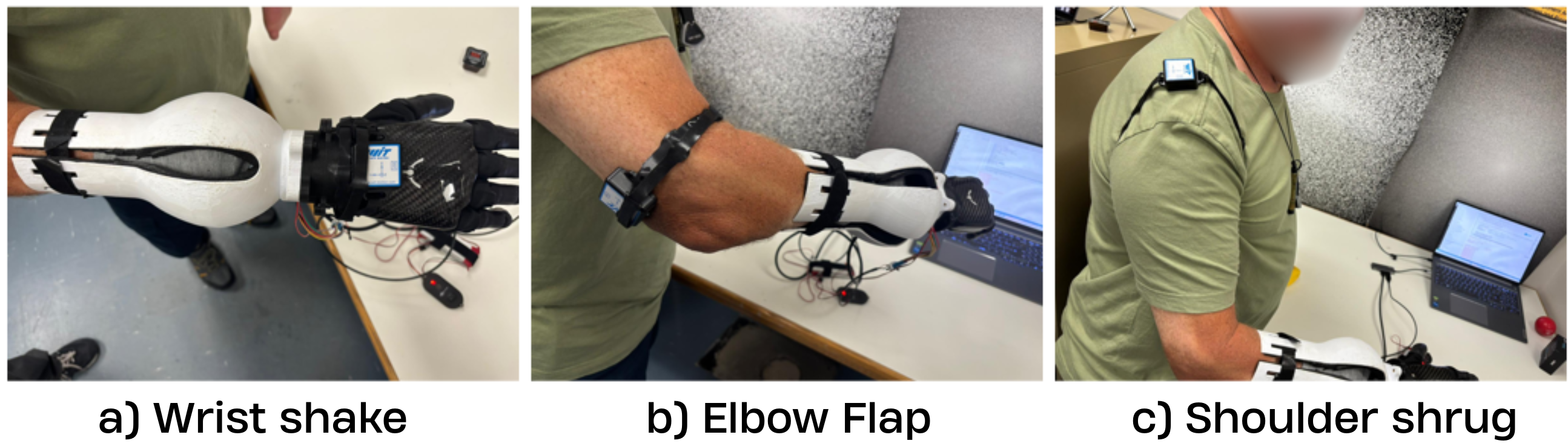}
        \caption{(a) Wrist-shake: IMU mounted on Prosthetic hand's wrist. (b) Elbow-flap: IMU tied above elbow. (c) Shoulder-shrug: IMU set up on top of shoulder}
        \label{fig:gesture_locations}
\end{figure}

These gestures were chosen based on multiple criteria: reliable detectability, ergonomic comfort, minimal fatigue, and low risk of accidental activation during daily activities.
The IMU was mounted at each gesture’s corresponding anatomical location: on the wrist for the wrist-shake gesture, above the elbow for the elbow-flap gesture, and on the top of the shoulder for the shoulder-shrug gesture.
Representative examples of these gesture configurations are shown in Fig.~\ref{fig:gesture_locations}.

\subsubsection{Signal Processing and Gesture Detection} 
\label{Gesture-detection}
The IMU device continuously streams tri-axial accelerometer and gyroscope data to the control computer, where a real-time Python algorithm performs motion detection. The \textit{IMU process} in the control framework manages asynchronous BLE communication to maintain a stable connection and continuously read and process live IMU data streams.
A sliding window of the most recent 40 samples (approximately 3.5-4.0~s of data at a sample rate of 10 Hz) is maintained~\cite{8}. 
Each updated window is analyzed by a dedicated \textit{gesture-detection} algorithm to identify the occurrence of a valid trigger gesture.
A defining characteristic of the trigger gestures in the IMU signal is the presence of pronounced fluctuations in acceleration, typically concentrated along a single dominant axis.
To detect these fluctuations, the gesture-detection algorithm employs the \texttt{scipy.signal.find\_peaks} function to locate local minima and maxima in the acceleration waves.
Any signal reading that crosses a predefined threshold is classified as an intense motion wave.
To adapt to signal variability and reduce false positives, dynamic thresholding strategy is employed, which derives thresholds $T_{\mathrm{peak}}$ and $T_{\mathrm{valley}}$ from the local average acceleration value \cite{16}, defined as bellow:  
\begin{equation}
    T_{\mathrm{peak}} = \Delta + S \cdot \Delta, \qquad
    T_{\mathrm{valley}} = \Delta - S \cdot \Delta ,
    \label{eq:thresholds}
\end{equation}
where $\Delta$ denotes the mean absolute acceleration magnitude over the sliding window and $S$ is a scalar multiplier empirically used to scale the adaptive threshold range. The threshold multiplier ($S$) was empirically fine-tuned for each gesture and set to 550 for the wrist-shake, 650 for the elbow-flap, and 450 for the shoulder-shrug gesture. 

\begin{figure}[htbp]
    \centering
    \includegraphics[width=0.48\textwidth]{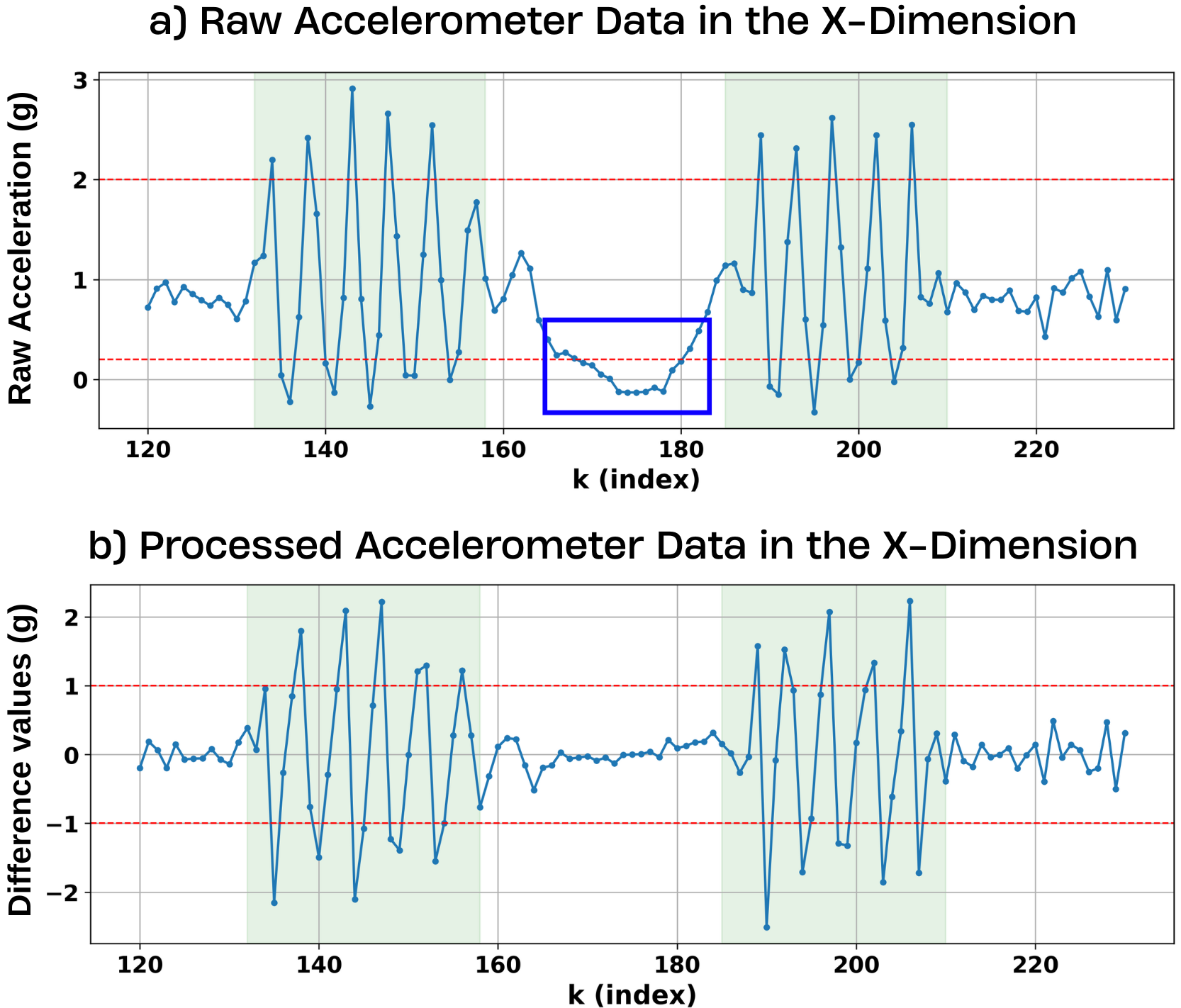}
    \caption{Illustration of IMU acceleration signal processing. (a) Raw X-axis acceleration showing target gesture segments (green) and non-gesture segments (blue box) that occasionally exceed dynamic thresholds (red dotted lines) due to drift. (b) Difference signal derived from the same data, where non-gesture fluctuations are suppressed and only intentional gestures cross the thresholds.}

    \label{fig:imu_processing}
\end{figure}

To suppress drift and emphasize rapid oscillations, the algorithm operated on the difference value in between the raw signal values ($d[k]$), computed as  
\begin{equation}
d[k] = a[k] - a[k-1], \quad k = 2, \ldots, N,
\end{equation}
where $a[k]$ is the raw acceleration magnitude at the $k$-th sample of the sliding window ($N=40$). This transformation highlights peak-valley alternations characteristic of deliberate gestures while reducing baseline drift (see Fig.~\ref{fig:imu_processing}).

Since the algorithm operated on difference values rather than raw acceleration signals, the mean acceleration during periods of low or steady activity could approach zero (e.g., 0.0004). In such cases, excessively low thresholds could lead to false-positive detections. To mitigate this, if the average acceleration in the last 40 samples (about 3.5 seconds) is below 0.001, then the thresholds were was assigned fixed baseline values, termed the zero-deviation thresholds, instead of relying on dynamic scaling. These parameters were empirically fine-tuned for each gesture and set to 0.35 for the wrist-shake, 0.35 for the elbow-flap, and 0.20 for the shoulder-shrug gesture.

\subsubsection{Gesture-to-Action Mapping} 
Once a valid gesture is detected, the manual control system determines the appropriate action based on the current hand state information received from the hand-control process.
If the hand is open, the system triggers a \textit{grasp} command; conversely, if the hand is closed, the same gesture initiates a \textit{release} (or relax) command\footnote{How the current hand state of grasping or relaxing is determined is described in detail in Section~\ref{hybrid}.}. 
This mapping ensures intuitive bidirectional control, allowing a single gesture to alternate between grasp and release actions depending on the prosthetic hand’s state.

The Manual Control architecture is illustrated in the bottom panel in Fig.~\ref{fig:sys_arch}.

\subsection{Hybrid Control: Integration of Manual Control with Autonomous System}
\label{hybrid}
The IMU-based manual control framework was integrated with the autonomous neural network (NN) controller to form a hybrid system that performs autonomous grasping and user-initiated release through IMU gestures.

A dedicated \textit{control-flow process} manages coordination between the autonomous and manual subsystems. It continuously monitors output from the NN, the gesture-detection process, and the hand-control process to determine the appropriate command for the prosthetic hand. Based on real-time hand-state information, if the hand is open (relaxing), the control-flow process enables the NN output to drive autonomous grasping. Conversely, if the hand is currently grasping an object, control authority transitions from the NN to the IMU-based manual system, allowing a voluntary user-triggered release.

Before transferring control from autonomous to manual operation, a \textit{secure grip} condition is verified to prevent weak grasps and ensure that the object is firmly held. This condition is determined by the tactile sensor data analyzed for an S-curve pattern, characterized by a rapid increase in grip force immediately before object contact is completed, followed by stabilization once the object can no longer be compressed, based on the most recent 30 readings obtained from the prosthetic hand’s fingertip force sensors.

Once a secure grip is confirmed, the control-flow process temporarily pauses the NN and activates the IMU-based manual control.
A valid user-trigger immediately issues a release command, after which NN control resumes.
This cooperative mechanism prevents premature release and ensures that objects are not dropped once a stable grasp has been achieved.
The Hybrid Control architecture is illustrated in Fig.~\ref{fig:sys_arch}.

\noindent
The described architecture supports three operational paradigms: autonomous, manual, and hybrid. each enabling a different balance of user agency and automation.

\section{Experiment}

A study was conducted to evaluate the performance of three trigger gestures and three control modes. The experiment was divided into two parts: the \textit{Gesture Evaluation Study} and the \textit{Control-Mode Evaluation Study}. 

\subsection{Subjects}

\begin{figure}
        \centering        \includegraphics[width=0.9\linewidth]{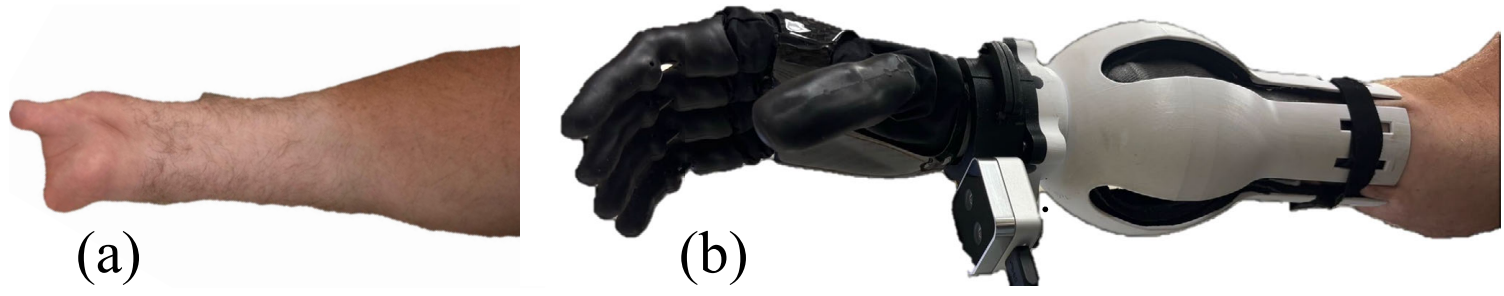}
        \caption{(a) Residual right forearm stump. (b) Participant wearing the prosthetic hand with the wrist-mounted camera attached.}
        \label{fig:stump_and_prosthetic}
\end{figure}

A total of fifteen participants were recruited for the gesture-evaluation study comprising fourteen able-bodied individuals (eight males and six females, all right-handed, ages 20-30 years) and one participant with a congenital bilateral limb difference (male, right-handed, 61 years). The individual exhibited underdeveloped hands distal to the wrist but retained sufficient skeletal structure to permit residual rotational mobility of the stump (see Fig.~\ref{fig:stump_and_prosthetic}), functionally analogous to wrist pronation.  As a result of this residual mobility, wrist pronation was included as an additional candidate gesture in the evaluation for the amputee participant, alongside the three standard gestures: wrist shake, elbow flap, and shoulder shrug.
A subset of twelve participants from the same cohort, consisting of eleven able-bodied individuals and one participant with a limb difference, subsequently participated in the control-mode evaluation study, in which the three operational modes (autonomous, hybrid, and manual) were assessed using the elbow-flap gesture as the trigger motion.

All participants involved in this study provided informed
consent before participating. The consent process was
conducted in accordance with the ethical standards outlined
by the institutional ethics committee. Participants were
informed about the purpose and nature of the study. They
were assured of confidentiality and the voluntary nature of
their participation. Additionally, participants were informed of
their right to withdraw from the study at any time without
consequence.

\subsection{Experiment Protocol}

The experimental protocol consisted of two studies: (1) Gesture Evaluation and (2) Control-Mode Evaluation. Both studies followed identical environmental and procedural settings to ensure comparability across conditions.

\subsubsection{Experiment tools}
\begin{figure}
    \centering
    \includegraphics[width=0.99\linewidth]{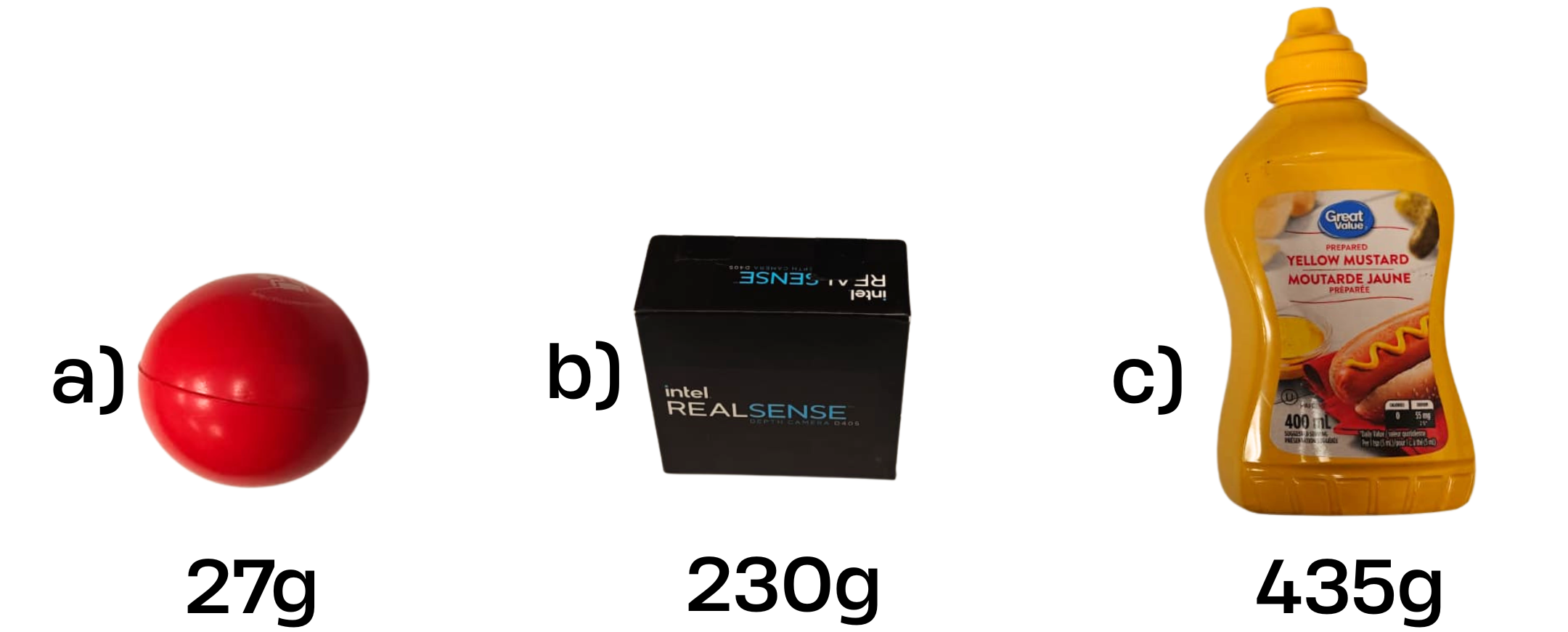}
    \caption{Three study objects used for testing control modes and trigger gestures: a red sponge ball (27 g), a filled black box (230 g), and a full mustard bottle (435 g)}
    \label{fig:objects}
\end{figure}

All tests were conducted using the PSYONIC Ability Hand worn on the participant’s right limb. A Bluetooth-enabled inertial measurement unit (IMU; WT901BLECL, WitMotion Inc.) transmitted real-time acceleration data to the control computer, where the gesture-detection algorithm operated continuously. Three standardized test objects were used throughout all experiments: a red sponge ball (27 g), a black cardboard box filled with screws (230 g), and a full mustard bottle (435 g), see Fig.\ref{fig:objects} for reference. 

\subsubsection{Trial Procedure}
Participants stood in front of a flat-topped table with one of the three standardized test objects placed within arm’s reach. For grasp trials, participants lifted the object using the prosthetic hand upon receiving a verbal cue and held it briefly in midair. For release trials, participants placed the held object back onto the table when instructed to release. Each participant performed ten trials per object (five grasp, five release), yielding 30 trials per each gesture and per each mode. A successful trial was defined as one in which the expected action (grasp or release) was completed on the first attempt. All experimental parameters-including object placement, IMU sampling rate, and algorithmic parameters, were kept identical across participants.

\subsubsection{Gesture Evaluation Specifics}
The Gesture Evaluation Study with fifteen participants (14 able-bodied, one with congenital limb difference) evaluated three candidate trigger gestures: wrist shake, elbow flap, and shoulder shrug. Each gesture was evaluated in manual mode, wherein the trigger gestures initiated both grasp and release actions of the prosthetic hand. The IMU was mounted at the anatomical site corresponding to each gesture: on the prosthetic wrist for the wrist-shake gesture, above the elbow on the upper arm for the elbow-flap gesture, and atop the shoulder for the shoulder-shrug gesture (see Fig.~\ref{fig:gesture_locations}). For the participant with a limb difference, wrist pronation was additionally evaluated in a supplementary study due to retained residual rotational mobility of the stump. The order of gesture candidates evaluation was counterbalanced to minimize learning effects or bias. 
To ensure consistent signal detection, gesture-specific threshold parameters were empirically tuned based on pilot trials and defined as follows:

\begin{itemize}
    \item \textbf{Elbow-flap:} principal axis: X, threshold multiplier  : 650, zero deviation: 0.35
    \item \textbf{Wrist-shake:} principal axis: X, threshold multiplier : 550, zero deviation: 0.35
    \item \textbf{Shoulder-shrug:} principal axis: Z, threshold multiplier : 450, zero deviation: 0.20
\end{itemize}

These parameters determine the adaptive thresholds used for peak-valley detection in the real-time motion detection algorithm described in section~\ref{Gesture-detection}. The \textit{principal axis} specifies the dominant direction of motion for each gesture (e.g., lateral motion along the X-axis for wrist-shake and vertical motion along the Z-axis for shoulder-shrug). The \textit{threshold multiplier} corresponds to the scalar ($S$) in~\eqref{eq:thresholds}, scaling the dynamic detection range for peak–valley identification. The \textit{zero-deviation parameter} establishes a baseline threshold that mitigates false activations during periods of minimal or steady acceleration.

Once the Gesture-evaluation phase was completed, participants completed subjective assessments. NASA-TLX questionnaires were used to evaluate workload across six dimensions: mental demand, physical demand, temporal demand, performance, effort, and frustration. Responses were recorded on a 0-10 scale, where 0 indicated “very low” and 10 indicated “very high.”  In addition, a custom gesture evaluation questionnaire captured qualitative user preferences. Participants ranked the three gestures according to physical ease, social comfort, IMU placement comfort, (concern regarding accidental triggers, and overall daily-use preference.

\subsubsection{Control Mode Evaluation Specifics}

A subset of twelve participants (11 able-bodied, 1 with limb difference) participated, to compare three control paradigms: Autonomous, Hybrid, and Manual.
\begin{itemize}
    \item In \textit{Autonomous Mode Evaluation}, the neural network (NN) executed both grasp and release automatically based on visual and tactile input; no IMU was used.
    \item In \textit{Hybrid Mode}, grasp actions were autonomous, while release was triggered by the user via the IMU using the elbow-flap gesture (the best-performing gesture from the previous phase).
    \item In \textit{Manual Mode}, both grasp and release were triggered by the IMU gesture alone, with the NN paused.
\end{itemize}

All other experimental conditions including: trial counts, object set, and order randomization, were identical to those in the gesture evaluation. Each participant completed 30 trials per mode (3 objects x 10 trials), resulting in a total of 360 trials all participants per mode. After completing this phase, participants again rated workload using the same NASA-TLX questionnaire and completed a custom Likert-scale questionnaire evaluating system reliability, responsiveness, perceived control, ease of use, and overall preference.

\subsection{Data Analysis and Evaluation}

All recorded trial data was processed to quantify performance and evaluate the comparative accuracy of each gesture and control mode. Each grasp or release trial was binary-coded, with a value of 1 assigned to successful completion and 0 to failure.

In total, 1350 trials were recorded during the Gesture-Evaluation study (450 per gesture) and 1080 during the Control-Mode study (360 per mode). However, because each release action depended on its preceding grasp, an unsuccessful grasp invalidated the paired release. Such dependent releases were logged as None and excluded from subsequent analysis, resulting in minor differences in valid-trial counts across gestures and modes.

For the Gesture Evaluation Study, the valid trial counts were:
\begin{itemize}
    \item \textbf{Elbow flap:} 432 total trials (221 grasp + 211 release)
    \item \textbf{Shoulder shrug}: 442 total trials (225 grasp + 217 release)
    \item \textbf{Wrist shake}: 428 total trials (225 grasp + 203 release)
\end{itemize}

For the Control-Mode Evaluation, the valid trial counts were:
\begin{itemize}
    \item \textbf{Autonomous mode}: 348 total trials (180 grasp + 168 release)
    \item \textbf{Hybrid mode:} 344 total trials (180 grasp + 164 release)
    \item \textbf{Manual mode}: 343 total trials (176 grasp + 167 release)
\end{itemize}

Eight trials (four grasp, four release) from the elbow-gesture/manual-mode evaluation (conducted as a shared test) were excluded due to improper IMU placement during one participant’s session, which resulted in unreliable motion data.

The raw data was imported into Python and processed using the pandas library. Each record contained participant ID (anonymous), trial number, gesture or mode label, object type, intended action (grasp or release), and binary outcome (1 for success, 0 for failure). 

Datasets were cleaned, merged, and grouped by gesture and by control mode. Descriptive statistics such as mean accuracy, variance, and standard deviation were computed, and the same aggregation was applied separately to grasp and release datasets to identify asymmetric performance trends.

To assess within-subject consistency, mean performance accuracies were computed per participant for each gesture and control mode, generating participant-averaged distributions. Statistical comparisons were performed using two-tailed paired t-tests  to assess differences among gestures and modes. A significance threshold of \( p < 0.05 \) was set. Reported p-values (using \textsc{scipy.stats.ttest\_rel}) in  Section~\ref{sec:Results} confirm whether differences were statistically significant.

\section{Results}
\label{sec:Results}
\subsection{Gesture Detection Accuracy}
\begin{figure}[ht]
    \centering
    \includegraphics[width=0.99\linewidth]{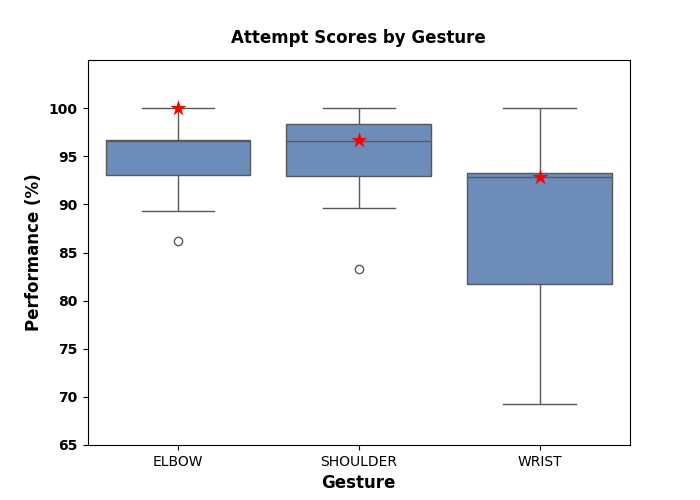}
    \caption{Average success rates of each Gesture across fourteen healthy participants and one participant with limb difference. Each participant’s data is an average of three objects (ball, bottle, box). The red five pointed star represents the person with limb difference.}
    \label{fig:Gesture_acc_BOX_PLOT}
\end{figure}

Gesture detection performance was evaluated across a total of 1302 valid trials. Figure \ref{fig:Gesture_acc_BOX_PLOT} presents the participant-wise average successful rates for each gesture in the form of comparative box plots.

The average successful rate calculated for wrist shake was 87.85\% out of 428 trials, shoulder shrug 95.02\% out of 442, and elbow flap achieved 95.37\% out of 432 trials. The corresponding accuracies are summarized in Table~\ref{tab:Gesture_all_accuracies}.  Paired t-tests confirmed that wrist shake successful rate was significantly lower than both shoulder shrug ($p = 0.0164$) and elbow flap ($p = 0.0274$), while no significant difference was found between shoulder shrug and elbow flap ($p = 0.7978$). 

\begin{table}[!t]
    \centering
    \caption{Trigger Gestures - Performance Accuracies}
    \label{tab:Gesture_all_accuracies}
    \setlength{\tabcolsep}{4pt} 
    \begin{tabular}{|c|c|c|c|}
        \hline
        \textbf{Gesture} & \textbf{Average Acc. (\%)} & \textbf{Grab Acc. (\%)} & \textbf{Release Acc. (\%)} \\ \hline
        \textbf{Elbow} & 95.37$\pm$21.03 & 95.48$\pm$20.83 & 95.26$\pm$21.30 \\ \hline
        \textbf{Shoulder} & 95.02$\pm$21.77 & 96.44$\pm$18.56 & 93.55$\pm$24.62 \\ \hline
        \textbf{Wrist}    & 87.85$\pm$32.70 & 90.22$\pm$29.77 & 85.22$\pm$35.58 \\ \hline
    \end{tabular}
\end{table}

The grasp versus release performance of each gesture is listed in Table~\ref{tab:Gesture_all_accuracies}. Paired t-tests revealed that for grasp successful rate, wrist shake performance is marginally lower than shoulder shrug ($p = 0.0529$) and not significantly different from elbow flap ($p = 0.1109$). No significant difference was observed between elbow flap and shoulder shrug ($p = 0.4332$). For release successful rate, wrist shake was significantly lower than both shoulder shrug ($p = 0.0239$) and elbow flap ($p = 0.0129$), while no difference was found between shoulder shrug and elbow flap ($p = 0.5268$). These results indicate that shoulder and elbow gestures achieved comparable grasp performance, whereas wrist-based triggering was less reliable, particularly during release actions.


\subsection{Control Mode Evaluation}
The control-mode evaluation comprised a total of 1115 valid trials. Fig.~\ref{fig:Mode_acc} illustrates participant-wise performance across modes using comparative box plots.

Autonomous mode achieved 85.91\% successful rate across 348 trials, hybrid mode 93.31\% across 344 trials, and manual mode 95.33\% across 343 trials. The corresponding accuracies are summarized in Table~\ref{tab:Mode_grasp_release}.  Hybrid mode marginally outperformed autonomous control ($p = 0.0701$), while manual mode showed a significant improvement over autonomous control ($p = 0.0422$). No significant difference was observed between hybrid and manual modes ($p = 0.3260$).

\begin{table}[h]
    \centering
    \caption{Control Modes - Performance Accuracies}
    \label{tab:Mode_grasp_release}
    \setlength{\tabcolsep}{4pt}
    \begin{tabular}{|c|c|c|c|}    
        \hline
        \textbf{Mode} & \textbf{Average Acc. (\%)} & \textbf{Grab Acc. (\%)} & \textbf{Release Acc. (\%)} \\ \hline
        \textbf{Autonomous} & 85.91$\pm$34.83 & 93.33$\pm$25.01 & 77.89$\pm$41.56 \\ \hline
        \textbf{Hybrid} & 93.31$\pm$25.01 & 91.11$\pm$28.54 & 95.73$\pm$20.28 \\ \hline
        \textbf{Manual} & 95.33$\pm$21.11 & 94.89$\pm$22.09 & 95.81$\pm$20.10 \\ \hline
    \end{tabular}
\end{table}

\begin{figure}[ht]
    \centering
    \includegraphics[width=0.99\linewidth]{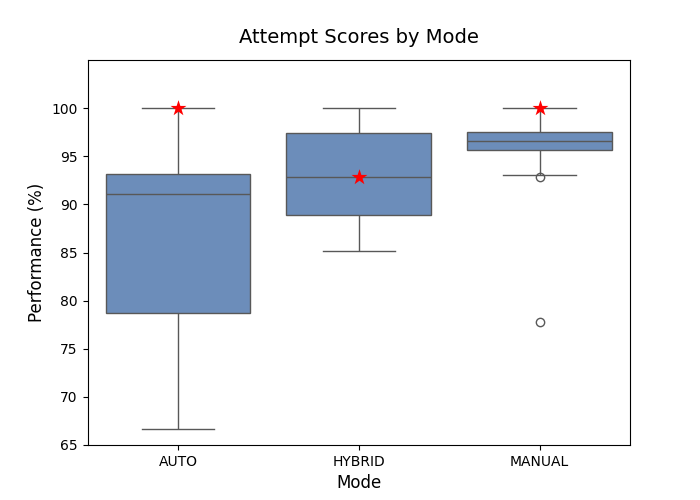}
    \label{fig:Mode_acc}
    \caption{Average success rates of each Mode across eleven healthy participants and one participant with limb difference. Each participant’s data is an average of three objects (ball, bottle, box). The red five pointed star represents the person with limb difference.}
    \label{fig:Mode_acc}
\end{figure}

Grasp and release successful rates for each mode are summarized in Table~\ref{tab:Mode_grasp_release}. No significant differences were observed in grasp successful rate across modes ($p > 0.05$), with $p = 0.4627$ for autonomous versus manual, $p = 0.3944$ for autonomous versus hybrid, and $p = 0.2064$ for manual versus hybrid comparisons. For release successful rate, both hybrid (95.5\%) and manual (95.4\%) modes significantly outperformed autonomous control (78.3\%), with $p = 0.0305$ and $p = 0.0334$, respectively. No significant difference was found between hybrid and manual modes ($p = 0.9208$).


\subsection{User Preference}
The subjective assessments show that the elbow-flap gesture was the most preferred, selected by 10 of the 15 participants (66\%), followed by the wrist-shake (4 participants out 15, 26.6\%) and the shoulder-shrug (1 participant, 6.7\%). Participants also provided detailed evaluations of each gesture across four qualitative dimensions: physical ease, social comfort, IMU placement comfort, and concern for accidental triggering, using a ranking scale where 1 indicated the most favorable and 3 the least favorable gesture. The results, summarized in Table~\ref{tab:gesture_rank}, show that the elbow-flap was ranked highest in physical ease ($1.40 \pm 0.51$) and social comfort ($1.80 \pm 0.68$), indicating it was perceived as both natural and socially comfortable to perform. The wrist-shake gesture ranked highest in IMU placement comfort ($1.40 \pm 0.74$) and was closely comparable to the elbow-flap in social comfort ($1.87 \pm 0.83$), but it was ranked poor in physical ease and accidental trigger concern. The shoulder-shrug gesture ranked best in the accidental trigger concern category, indicating a low perceived chance of false activations, although rankings for this category were mostly similar across all gestures. Overall, the shoulder-shrug gesture ranked poorest across the remaining categories, reflecting participant feedback that it felt physically straining and less practical for frequent use. These findings reinforce the quantitative results, confirming that the elbow-flap was the most ergonomic and intuitive gesture overall.

For control paradigms, the hybrid and autonomous modes received equal preference (4.5 of 12 votes each, 37.5\%), whereas the manual mode was least favored (3 of 12 votes, 25\%). In cases where participants could not choose between two modes, their votes were evenly split between the selected options. Participants also rated each mode across four subjective criteria: system reliability, responsiveness, perceived control, and ease of use, using a five-point Likert scale (1 = strongly disagree, 5 = strongly agree). The resulting mean ratings are summarized in Table~\ref{tab:mode_custom}. Manual mode achieved the highest overall average score ($3.75/5$), reflecting greater perceived reliability ($4.17 \pm 0.83$) and control ($4.17 \pm 0.94$) due to its high user input and consistent performance accuracy, while hybrid mode scored highest in responsiveness ($3.67 \pm 0.65$) and ease of use ($4.00 \pm 1.04$), since the hybrid maintained prompt reactions with reduced effort. The autonomous mode received the lowest ratings across all categories.

\begin{table}[!t]
\centering
\setlength{\tabcolsep}{2pt}
\caption{Average Rank ($N=15$) for Gesture Evaluation Questions (1 = Most Favorable, 3 = Least Favorable)}
\begin{tabular}{|c|c|c|c|}
\hline
\textbf{Question} & \textbf{Wrist Shake} & \textbf{Elbow Flap} & \textbf{Shoulder Shrug} \\
\hline
Physical Ease & $2.13 \pm 0.83$ & \textbf{1.40} $\pm$ 0.51 & $2.47 \pm 0.74$ \\
Social Comfort & $1.87 \pm 0.83$ & \textbf{1.80} $\pm 0.68$ & $2.40 \pm 0.91$ \\
IMU Placement Comfort & \textbf{1.40} $\pm 0.74$ & $2.07 \pm 0.70$ & $2.53 \pm 0.64$ \\
Accidental Trigger Concern & $2.13 \pm 0.76$ & $1.93 \pm 0.69$ & \textbf{1.73} $\pm 0.70$ \\
\hline
\end{tabular}
\label{tab:gesture_rank}
\end{table}

\begin{table}[t]
\centering
\caption{Average Subjective Ratings (1-5) for Each Control Mode (N=12)}
\label{tab:mode_custom}
\setlength{\tabcolsep}{6pt}
\begin{tabular}{|c|c|c|c|}
\hline
\textbf{Criterion} & \textbf{Autonomous} & \textbf{Hybrid} & \textbf{Manual} \\
\hline
System Reliability      & 2.58 $\pm$ 1.08 & 3.58 $\pm$ 0.67 & \textbf{4.17} $\pm$ 0.83 \\
System Responsiveness   & 3.33 $\pm$ 1.37 & \textbf{3.67} $\pm$ 0.65 & 3.25 $\pm$ 1.22 \\
Feeling of Control      & 2.25 $\pm$ 1.22 & 3.25 $\pm$ 0.87 & \textbf{4.17} $\pm$ 0.94 \\
Ease of Use             & 3.33 $\pm$ 1.44 & \textbf{4.00} $\pm$ 1.04 & 3.42 $\pm$ 1.24 \\ \hline
\textbf{Overall Average Score}    & 2.8 / 5 & 3.63 / 5 & \textbf{3.75 / 5} \\
\hline
\end{tabular}
\end{table}

\subsection{Case Study: Participant with Limb Difference}
A participant with a congenital bilateral limb difference was included in the evaluation. In addition to the three standard gestures, wrist pronation was tested as an alternative trigger. The participant identified wrist pronation as the most preferred gesture overall. Among the standard gestures, elbow flap was favored. Participant reported that wrist shake required exaggerated motion to be reliably detected, while shoulder shrug was described as physically uncomfortable.

For control modes, the participant reported the highest comfort with manual mode due to its predictability but indicated that autonomous mode would be preferable for daily use if reliability were improved.

\subsection{Workload Evaluation (NASA-TLX)}
\begin{figure*}[!t]
    \centering    \includegraphics[width=\textwidth]
    {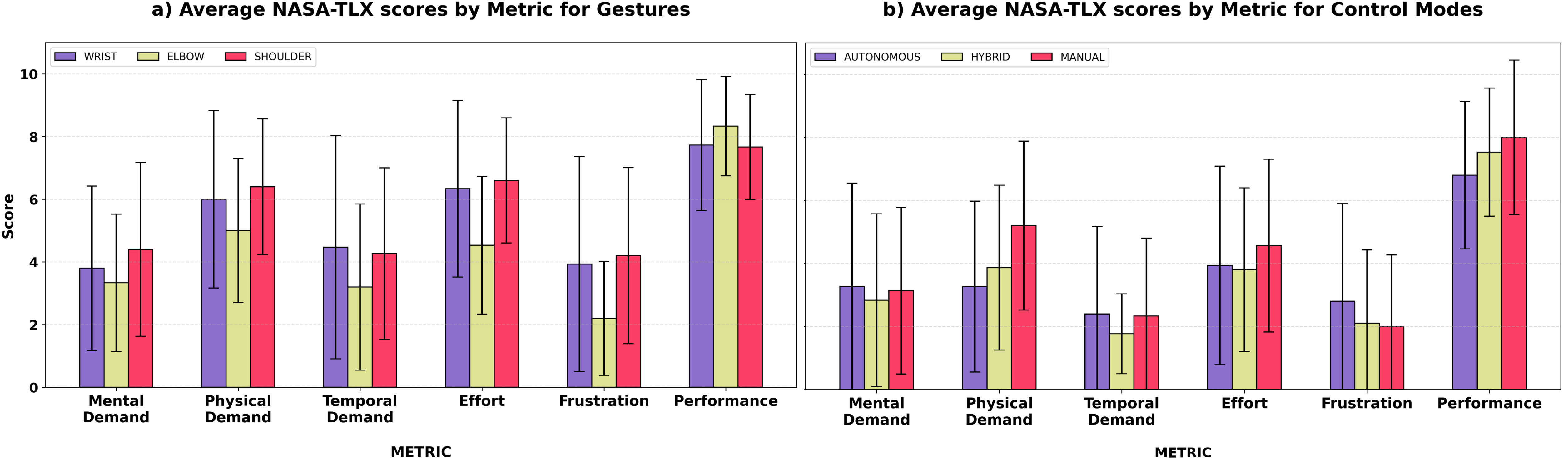}
    \caption{(a) Average NASA-TLX scores across participants for each trigger gesture. Each colour represents one gesture. (b) Average NASA-TLX scores across participants for each control mode. Each colour represents one mode.}
    \label{fig:Nasa_results}
\end{figure*}

The NASA Task Load Index (NASA-TLX) results (see Fig.~\ref{fig:Nasa_results}) indicate that the elbow-flap gesture yielded the most favorable overall workload profile. It required the least effort (4.53/10), lowest frustration (2.20/10), and lowest mental demand (3.33/10), while achieving the highest perceived performance (8.33/10). It also demonstrated the lowest physical (5.00/10) and temporal (3.20/10) demands. The wrist-shake gesture ranked second, showing moderate effort (6.33/10), frustration (3.93/10), mental demand (3.80/10), performance (7.73/10), and physical demand (6.00/10), but exhibited the highest temporal demand (4.47/10). The shoulder-shrug gesture showed the poorest overall workload, with the highest effort (6.60/10), frustration (4.20/10), mental demand (4.40/10), physical demand (6.40/10), and the lowest performance (7.67/10). Its temporal demand (4.27/10) was close to wrist-shake's temporal demand. 

Across control modes, hybrid achieved the best overall workload, with the least effort (3.80/10), lowest mental demand (3.00/10), and lowest temporal demand (1.77/10). Manual mode yielded the lowest frustration score (2.00/10) and highest perceived performance (8.00/10) but exhibited higher physical (5.20/10) and effort (4.57/10) demands. Autonomous mode showed the lowest physical demand (3.27/10) but lower perceived performance (6.80/10) compared with the other modes. Overall, the results suggest that the hybrid mode provided the best balance of low workload and high usability among the three control paradigms.

\section{Discussion} 


This study demonstrates how IMU-based voluntary triggers can effectively restore user agency in shared-autonomous prosthetic hand control, particularly in scenarios where autonomous behavior diverges from user intent. Beyond confirming technical feasibility, the findings elucidate key dynamics between objective performance metrics and subjective user experience.

Gesture evaluation results highlight the relationship between ergonomic comfort and user preference. Although both the elbow-flap and shoulder-shrug gestures achieved comparably high detection accuracies, participants overwhelmingly favored the elbow-flap, which received high ratings across all NASA–TLX dimensions as well as superior rankings in \textit{physical ease} and \textit{social comfort}. These results suggest that the elbow-flap combines biomechanical efficiency with social acceptability, making it an intuitive and sustainable long-term trigger motion. The shoulder-shrug gesture was the least preferred, scoring lowest across all NASA-TLX workload dimensions and most of the custom evaluation rankings, indicating that it was physically and ergonomically less suitable for frequent use. Participants described it as unnatural and fatiguing; one noted, “The shoulder gesture involves vertical rather than lateral movement, making it the most unnatural and physically straining.” Although the wrist-shake gesture achieved moderate performance, participants reported fatigue during prolonged use, particularly with heavier objects. They nevertheless favored the IMU’s wrist placement, describing it as natural and familiar—“it feels like wearing a watch.” These results underscore that ergonomic comfort and motion naturalness have greater impact on user acceptance than performance accuracy alone.

Gesture evaluation results from the participant with limb difference provide preliminary insights into the system’s real-world applicability. The strong preference for wrist pronation highlights the importance of tailoring gesture selection to an individual’s residual mobility. The participant noted that, although the elbow-flap gesture offered greater stability and bilateral usability, wrist pronation was ultimately preferred for its quick execution and minimal movement compared to the three standard gestures. These findings suggest that designing triggers requiring smaller, faster, and less effortful motions may enhance convenience and accessibility for users with varying limb conditions.

Control-mode evaluation results also show that objective performance does not necessarily align with user preference. Although the manual mode achieved the highest task accuracy and scored strongly in perceived \textit{reliability}, \textit{control}, and the NASA–TLX \textit{performance} dimension, it was rated least favorable due to the continuous effort required for operation. In contrast, the autonomous mode received a relatively high preference rating despite exhibiting lower performance accuracy, indicating that users prioritize reduced workload and convenience over precision and reliability. Hybrid mode demonstrated the most balanced performance overall, achieving accuracy comparable to the manual mode and subjective preference similar to the autonomous mode. It also attained high or intermediate scores across most NASA–TLX dimensions. These results indicate that users favor systems that reduce physical and cognitive effort while preserving agency at critical decision points; a balance that can strengthen user trust and long-term acceptance. However, participants reported certain drawbacks about Hybrid control. Some found the autonomous grasp adjustments unexpected or unsettling, whereas others considered the repeated gesture inputs for release actions still burdensome. One participant also noted that the mismatch between automated grasping and manual release occasionally caused confusion.

An additional control method called Override Mode was implemented to explore concurrent control between the IMU gesture detector and the neural network. In this configuration, user gestures immediately override ongoing autonomous actions to execute the opposite command (if hand is autonomously opening, doing the gesture will close the hand and vise versa). Although this mode was not formally evaluated, it was informally tested by a participant with a limb difference, who reported it to be intuitive, responsive, and preferable to the other three control modes. These preliminary observations suggest that concurrent override control could enhance user trust and safety by preserving autonomy while maintaining immediate user control. This finding underscores a preference for flexible systems that retain autonomy under explicit user authority. 

The balanced performance of the hybrid mode highlights the potential of mode switching based on task demands. For routine activities, hybrid or override modes may provide an optimal compromise between autonomy and manual control, whereas manual operation may remain valuable for precision-critical or safety-sensitive tasks. This adaptive control paradigm could be extended to context-aware systems that dynamically adjust the degree of autonomy based on user behavior or environment.

Finally, the clear differences between able-bodied and limb-different participants underscore the importance of inclusive evaluation. While able-bodied participants prioritized comfort and reduced fatigue, the participant with limb difference emphasized reliability and predictability as primary concerns. These differences suggest that prosthetic interfaces must be adaptable not only across tasks but also across user populations, recognizing varied physical capabilities and lived experiences.

Another noteworthy aspect concerns the latency differences observed among the control modes. Although informal observations indicated that autonomous control completed grasp–release cycles in approximately 2-3~s, hybrid control introduced about a 1-2~s delay due to the IMU-triggered release, and manual control extended total cycle time to roughly 5-6~s. These differences may have influenced user preferences, as faster response times in the autonomous mode were often perceived as smoother and less effortful. However, since timing data were not formally recorded, latency effects remain speculative and warrant quantitative evaluation in future work.

Another consideration involves user bias. Several participants expressed fascination with the autonomous mode’s human-like behavior, despite performing more accurately in hybrid or manual control. This novelty effect may have inflated preference ratings for the autonomous system, especially among first-time users unfamiliar with intelligent prosthetic behavior. Longitudinal studies with repeated exposure would be valuable to separate genuine usability improvements from initial novelty bias.

\section{Limitations and Future Work} 
While IMU-based control offers advantages over traditional sEMG (reduced physical strain, improved generalizability, and robustness to sweat and fatigue) it is currently limited to binary actions (grasp and release). By contrast, sEMG systems can discriminate among multiple grip types \cite{14}, \cite{15}. Expanding IMU-based control to support richer functionality is therefore a key direction. Possible approaches include mapping distinct gestures to specific grip types (e.g., elbow flap for spherical grip, wrist rotation for pinch grip), or applying advanced machine learning to infer user intent from subtle motion cues \cite{11}\cite{6}. Techniques such as multi-axis signal fusion \cite{10}, time-series gesture classification \cite{8}, and context-aware control logic may further enhance versatility. 

Although qualitative observations suggested noticeable timing latency among the control modes, future work will mitigate uncertainty and improve user awareness of system state. Feedback modalities such as LED indicators and vibration cues will be integrated to signal successful gesture detection, command execution, or grasp completion. 

Finally, while the present study validated IMU-based control with able-bodied participants and one individual with limb difference, broader testing is required to assess generalization to diverse user groups, including individuals with varying amputation levels or limited residual mobility. Real-world testing in unstructured environments will be essential to evaluate robustness against incidental motion, fatigue, and environmental interference. Moreover, the observed potential of the IMU-based override mode suggests the need for formal evaluation with larger participant cohorts. Such investigations will help refine trigger algorithms to minimize false positives, reduce response latency, and ensure consistent reliability under practical daily-use conditions.

\section{Conclusion} 
This study presents an IMU-based voluntary trigger system as a robust and intuitive interface for augmenting autonomous prosthetic hand control. By recognizing deliberate upper-limb motion gestures, the system enables users to initiate or override prosthetic actions in real time, addressing a critical limitation of fully autonomous systems.
Among the tested gestures, the elbow flap emerged as the most reliable and comfortable, while the hybrid control mode struck the optimal balance between automation and user agency. The proposed real-time gesture detection algorithm demonstrated high accuracy across users and scenarios, validating the practicality of IMU-based control in assistive technologies.
These findings underscore the potential of integrating IMU-triggered manual override into future prosthetic systems to enhance usability, safety, and adaptability. Further research should explore multi-gesture input schemes, real-world deployment with individuals with limb differences, and advanced signal processing techniques to support a broader range of functional tasks and grip types.

\section{ACKNOWLEDGMENT}
The authors would like to thank Ann Shigeishi and the Dr. Leonard A. Miller Center team for invaluable assistance with amputee participant recruitment and prosthetic technical support.

\nocite{*}

\bibliographystyle{IEEEtran}
\bibliography{references}

\begin{thebibliography}{10}
\providecommand{\url}[1]{#1}
\csname url@samestyle\endcsname
\providecommand{\newblock}{\relax}
\providecommand{\bibinfo}[2]{#2}
\providecommand{\BIBentrySTDinterwordspacing}{\spaceskip=0pt\relax}
\providecommand{\BIBentryALTinterwordstretchfactor}{4}
\providecommand{\BIBentryALTinterwordspacing}{\spaceskip=\fontdimen2\font plus
\BIBentryALTinterwordstretchfactor\fontdimen3\font minus \fontdimen4\font\relax}
\providecommand{\BIBforeignlanguage}[2]{{%
\expandafter\ifx\csname l@#1\endcsname\relax
\typeout{** WARNING: IEEEtran.bst: No hyphenation pattern has been}%
\typeout{** loaded for the language `#1'. Using the pattern for}%
\typeout{** the default language instead.}%
\else
\language=\csname l@#1\endcsname
\fi
#2}}
\providecommand{\BIBdecl}{\relax}
\BIBdecl

\bibitem{18}
\BIBentryALTinterwordspacing
M.~Savage \emph{et~al.}, ``Product narrative: Digital assistive technology. a market landscape and strategic approach to increasing access to digital assistive technology in low- and middle-income countries,'' Clinton Health Access Initiative (CHAI), London, U.K., Tech. Rep., 2020, tech. Rep. [Online]. Available: \url{https://www.at2030.org/product-narrative:-digital-assistive-technology/}
\BIBentrySTDinterwordspacing

\bibitem{26}
C.~L. McDonald, S.~Westcott-McCoy, M.~R. Weaver, J.~Haagsma, and D.~Kartin, ``Global prevalence of traumatic non-fatal limb amputation,'' \emph{Prosthetics and Orthotics International}, vol.~45, no.~2, pp. 105--114, 2021.

\bibitem{27}
J.~Frölke, G.~Rommers, A.~de~Boer, T.~Groenveld, and R.~Leijendekkers, ``Epidemiology of limb amputations and prosthetic use,'' \emph{Archives of Physical Medicine and Rehabilitation}, vol. 105, no.~2, pp. 280--286, 2024.

\bibitem{8}
\BIBentryALTinterwordspacing
F.~Schweitzer and A.~Campeau-Lecours, ``Imu-based hand gesture interface implementing a sequence-matching algorithm for the control of assistive technologies,'' \emph{Signals}, vol.~2, pp. 729--753, 2021. [Online]. Available: \url{https://doi.org/10.3390/signals2040043}
\BIBentrySTDinterwordspacing

\bibitem{19}
\BIBentryALTinterwordspacing
M.~B.~I. Reaz, M.~S. Hussain, and F.~Mohd-Yasin, ``Techniques of emg signal analysis: detection, processing, classification and applications,'' \emph{Biol. Proced. Online}, vol.~8, pp. 11--35, 2006. [Online]. Available: \url{https://doi.org/10.1251/bpo115}
\BIBentrySTDinterwordspacing

\bibitem{10}
\BIBentryALTinterwordspacing
A.~Krasoulis, I.~Kyranou, M.~S. Erden, K.~Nazarpour, and S.~Vijayakumar, ``Improved prosthetic hand control with concurrent use of myoelectric and inertial measurements,'' \emph{Journal of NeuroEngineering and Rehabilitation}, vol.~14, p.~71, 2017. [Online]. Available: \url{https://doi.org/10.1186/s12984-017-0284-4}
\BIBentrySTDinterwordspacing

\bibitem{4}
\BIBentryALTinterwordspacing
X.~Li, O.~W. Samuel, X.~Zhang, H.~Wang, P.~Fang, and G.~Li, ``A motion‐classification strategy based on semg‐eeg signal combination for upper‐limb amputees,'' \emph{Journal of NeuroEngineering and Rehabilitation}, vol.~14, no.~2, 2017. [Online]. Available: \url{https://doi.org/10.1186/s12984-016-0212-z}
\BIBentrySTDinterwordspacing

\bibitem{13}
\BIBentryALTinterwordspacing
P.~L. Bishay, J.~Wilgus, R.~Chen, D.~Valenzuela, V.~Medina, C.~Tan, T.~Ittner, M.~Caldera, C.~Rubalcava, S.~Safarian, G.~Funes~Alfaro, A.~Gonzalez-Martinez, M.~Gosparini, J.~Fuentes-Perez, A.~Lima, J.~Villalobos, and A.~Solis, ``Controlling a below-the-elbow prosthetic arm using the infinity foot controller,'' \emph{Prosthesis}, vol.~5, no.~4, pp. 1206--1231, Nov. 2023. [Online]. Available: \url{https://doi.org/10.3390/prosthesis5040084}
\BIBentrySTDinterwordspacing

\bibitem{1}
B.-Y.~Y. Jian-Wei~Cui, Han~Du and P.-D. Lu, ``Recognition of upper limb action intention based on imu,'' \emph{Sensors}, p.~2, 2022.

\bibitem{21}
R.~Volkmar \emph{et~al.}, ``Improving bimanual interaction with a prosthesis using semi-autonomous control,'' \emph{Journal of NeuroEngineering and Rehabilitation}, vol.~16, no.~1, p. 140, 2019.

\bibitem{22}
X.~Shi, W.~Guo, W.~Xu, Z.~Yang, and X.~Sheng, ``Semi-autonomous grasping control of prosthetic hand and wrist based on motion prior field,'' \emph{IEEE Robotics and Automation Letters}, vol.~9, no.~6, pp. 5950--5957, 2024.

\bibitem{23}
Y.~He, R.~Kubozono, O.~Fukuda, N.~Yamaguchi, and H.~Okumura, ``Vision-based assistance for myoelectric hand control,'' \emph{IEEE Access}, vol.~8, pp. 201\,956--201\,965, 2020.

\bibitem{24}
D.~P. McMullen, G.~Hotson, K.~D. Katyal, B.~A. Wester, M.~S. Fifer, T.~G. McGee, A.~Harris, M.~S. Johannes, R.~J. Vogelstein, A.~D. Ravitz \emph{et~al.}, ``Demonstration of a semi-autonomous hybrid brain--machine interface using human intracranial eeg, eye tracking, and computer vision to control a robotic upper limb prosthetic,'' \emph{IEEE Transactions on Neural Systems and Rehabilitation Engineering}, vol.~22, no.~4, pp. 784--796, 2013.

\bibitem{25}
N.~Ng, ``Development of a semantic model and synthetic dataset for multi-grasp affordance detection for application to vision-based upper-limb prosthetic grasping,'' Master's thesis, University of Waterloo, 2024.

\bibitem{5}
\BIBentryALTinterwordspacing
S.~Wang, J.~Zheng, Z.~Huang, X.~Zhang, V.~Prado~da Fonseca, B.~Zheng, and X.~Jiang, ``Integrating computer vision to prosthetic hand control with semg: Preliminary results in grasp classification,'' \emph{Frontiers in Robotics and AI}, vol.~9, 2022. [Online]. Available: \url{https://doi.org/10.3389/frobt.2022.948238}
\BIBentrySTDinterwordspacing

\bibitem{Shared-survery}
W.~Guo, W.~Xu, Y.~Zhao, X.~Shi, X.~Sheng, and X.~Zhu, ``Toward human-in-the-loop shared control for upper-limb prostheses: a systematic analysis of state-of-the-art technologies.'' \emph{IEEE transactions on Medical Robotics and Bionics}, vol.~5, no.~3, pp. 563--579, 2024.

\bibitem{20}
R.~J. Greene, D.~Kim, R.~Kaliki, P.~Kazanzides, and N.~Thakor, ``Shared control of upper limb prosthesis for improved robustness and usability,'' in \emph{BioRob 2022 - 9th IEEE RAS/EMBS International Conference on Biomedical Robotics and Biomechatronics}.\hskip 1em plus 0.5em minus 0.4em\relax IEEE Computer Society, August 2022.

\bibitem{kaijie}
K.~Shi, W.~Lu, H.~Zhao, V.~P. da~Fonseca, T.~Zou, and X.~Jiang, ``Toward biosignals-free autonomous prosthetic hand control via imitation learning,'' \emph{arXiv preprint arXiv:2506.08795}, Jun. 2025.

\bibitem{7}
\BIBentryALTinterwordspacing
G.~Ghazaei, A.~Alameer, P.~Degenaar, G.~Morgan, and K.~Nazarpour, ``Deep learning-based artificial vision for grasp classification in myoelectric hands,'' \emph{Journal of Neural Engineering}, vol.~14, no.~3, pp. 13--15, 2017. [Online]. Available: \url{https://doi.org/10.1088/1741-2552/aa6802}
\BIBentrySTDinterwordspacing

\bibitem{12}
\BIBentryALTinterwordspacing
F.~Rizzoglio, C.~Pierella, D.~De~Santis, F.~Mussa-Ivaldi, and M.~Casadio, ``A hybrid body-machine interface integrating signals from muscles and motions,'' \emph{Journal of Neural Engineering}, vol.~17, no.~4, p. 046004, Jul. 2020, pMID: 32521522. [Online]. Available: \url{https://doi.org/10.1088/1741-2552/ab9b6c}
\BIBentrySTDinterwordspacing

\bibitem{11}
\BIBentryALTinterwordspacing
E.~Valarezo~Añazco, S.-J. Han, K.~Kim, P.~Lopez, T.-S. Kim, and S.~Lee, ``Hand gesture recognition using single patchable six‐axis inertial measurement unit via recurrent neural networks,'' \emph{Sensors}, vol.~21, p. 1404, 2021. [Online]. Available: \url{https://doi.org/10.3390/s21041404}
\BIBentrySTDinterwordspacing

\bibitem{16}
M.~Pielka, P.~Janik, M.~A. Janik, and Z.~Wróbel, ``{Adaptive Data Transmission Algorithm for the System of Inertial Sensors for Hand Movement Acquisition},'' \emph{Sensors}, vol.~22, no.~24, p. 9866, Dec. 2022, pMID: 36560234; PMCID: PMC9781473.

\bibitem{14}
\BIBentryALTinterwordspacing
X.~Jiang, L.-K. Merhi, Z.~G. Xiao, and C.~Menon, ``Exploration of force myography and surface electromyography in hand gesture classification,'' \emph{Medical Engineering \& Physics}, vol.~41, pp. 63--73, 2017. [Online]. Available: \url{https://doi.org/10.1016/j.medengphy.2017.01.015}
\BIBentrySTDinterwordspacing

\bibitem{15}
\BIBentryALTinterwordspacing
S.~Wang, J.~Zheng, B.~Zheng, and X.~Jiang, ``Phase-based grasp classification for prosthetic hand control using semg,'' \emph{Biosensors}, vol.~12, no.~2, p.~57, 2022. [Online]. Available: \url{https://doi.org/10.3390/bios12020057}
\BIBentrySTDinterwordspacing

\bibitem{6}
M.~S.~B. Hossain, Z.~Guo, and H.~Choi, ``Estimation of lower extremity joint moments and 3d ground reaction forces using imu sensors in multiple walking conditions: A deep learning approach,'' \emph{IEEE Journal of Biomedical and Health Informatics}, vol.~27, no.~6, pp. 2829--2840, Jun. 2023.

\bibitem{2}
J.~G. Coll\'{i}~Alfaro, ``Implementation of user-independent hand gesture recognition classification models using imu and emg-based sensor fusion techniques,'' Master's thesis, University of Western Ontario, Electronic Thesis and Dissertation Repository, 2019.

\bibitem{3}
A.~Palkowski and G.~Redlarski, ``Basic hand gestures classification based on surface electromyography,'' \emph{Computational and Mathematical Methods in Medicine}, vol. 2016, pp. 1--3, 2016, article ID 6481282.

\bibitem{17}
T.~Feix, J.~Romero, H.-B. Schmiedmayer, A.~M. Dollar, and D.~Kragic, ``{The GRASP Taxonomy of Human Grasp Types},'' \emph{IEEE Transactions on Human–Machine Systems}, vol.~46, no.~1, pp. 66--77, Feb. 2016.

\bibitem{jiang2020estimating}
\BIBentryALTinterwordspacing
X.~Jiang, C.~Napier, B.~Hannigan, J.~J. Eng, and C.~Menon, ``{Estimating Vertical Ground Reaction Force during Walking Using a Single Inertial Sensor},'' \emph{Sensors}, vol.~20, no.~15, p. 4345, Aug. 2020. [Online]. Available: \url{https://doi.org/10.3390/s20154345}
\BIBentrySTDinterwordspacing

\bibitem{asfour2022feature}
\BIBentryALTinterwordspacing
M.~Asfour, C.~Menon, and X.~Jiang, ``{Feature–Classifier Pairing Compatibility for sEMG Signals in Hand Gesture Recognition under Joint Effects of Processing Procedures},'' \emph{Bioengineering}, vol.~9, no.~11, p. 634, 2022. [Online]. Available: \url{https://doi.org/10.3390/bioengineering9110634}
\BIBentrySTDinterwordspacing

\bibitem{alencar2024embedded}
M.~Alencar, R.~Barreto, H.~Oliveira, and E.~Souto, ``Embedded restricted boltzmann machine approach for adjustments of repetitive physical activities using imu data,'' \emph{IEEE Embedded Systems Letters}, vol.~16, no.~2, pp. 102--105, Jun. 2024.

\bibitem{stanzani2020measurement}
R.~Stanzani, P.~Dondero, A.~Mantero, and M.~Testa, ``Measurement accuracy of an upper limb tracking system based on two hillcrest labs bno080 imu sensors: An environmental assessment,'' \emph{IEEE Sensors Journal}, vol.~20, no.~17, pp. 10\,267--10\,274, Sep. 2020.

\bibitem{galviati2023imu}
R.~Galviati, N.~Boccardo, M.~Canepa, D.~Di~Domenico, A.~Marinelli, C.~A. Frigo, M.~Laffranchi, and L.~de~Michieli, ``Imu sensors measurements towards the development of novel prosthetic arm control strategies,'' in \emph{Proceedings of the IEEE International Conference on Rehabilitation Robotics (ICORR)}, Sep. 2023, pp. 1--6, pMID: 37941218.

\bibitem{kyranou2016real}
I.~Kyranou, A.~Krasoulis, M.~S. Erden, K.~Nazarpour, and S.~Vijayakumar, ``Real-time classification of multi-modal sensory data for prosthetic hand control,'' in \emph{2016 6th IEEE International Conference on Biomedical Robotics and Biomechatronics (BioRob)}, Singapore, 2016, pp. 536--541.

\bibitem{odeyemi2023exploring}
J.~Odeyemi, ``{Exploring the Potential of Computer Vision and Machine Learning in Enhancing the Functionality of an EMG-Controlled Prosthetic Hand},'' Master’s thesis, Memorial University of Newfoundland, St. John’s, NL, Canada, Aug. 2023.

\bibitem{Mohammadreza2025}
E.~Mohammadreza, V.~P.~D. Fonseca, and X.~Jiang, ``Investigating the impact of training protocols on myoelectric pattern recognition control in upper-limb amputees,'' \emph{IEEE Transactions on Neural Systems and Rehabilitation Engineering}, vol.~33, pp. 1338--1348, 2025, epub 2025 Apr 14.

\end{thebibliography}

\end{document}